\documentclass{IOS-Book-Article}

\usepackage{mathptmx}
\usepackage{alltt}
\usepackage{listings}
\usepackage{graphicx}
\usepackage{lscape}
\usepackage{rotating}
\usepackage{pdflscape}
\usepackage{adjustbox}
\usepackage{tabularx}
\usepackage{multirow}
\usepackage[numbers]{natbib}
\usepackage{algorithmic}
\usepackage[ruled,vlined]{algorithm2e}
\usepackage{url}
\usepackage{amsmath,amssymb,amsfonts}
\usepackage{float}
\usepackage{tikz}
\usepackage{color}
\usepackage{multicol}
\usepackage{enumitem}
\usepackage{pgfplots}
\usepackage{pgfplotstable}
\pgfplotsset{compat=1.17}
\usepackage{subcaption}
\usepackage{diagbox} 
\usepackage{hhline}
\usepackage{comment}

\usepackage[para]{footmisc}

\usepackage{amsthm}

\pgfplotscreateplotcyclelist{mycolorlist}{%
blue,every mark/.append style={fill=blue!80!black},mark=*\\%
red,every mark/.append style={fill=red!80!black},mark=square*\\%
brown!60!black,every mark/.append style={fill=brown!80!black},mark=otimes*\\%
black,mark=star\\%
blue,every mark/.append style={fill=blue!80!black},mark=diamond*\\%
red,densely dashed,every mark/.append style={solid,fill=red!80!black},mark=*\\%
brown!60!black,densely dashed,every mark/.append style={
solid,fill=brown!80!black},mark=square*\\%
black,densely dashed,every mark/.append style={solid,fill=gray},mark=otimes*\\%
blue,densely dashed,mark=star,every mark/.append style=solid\\%
red,densely dashed,every mark/.append style={solid,fill=red!80!black},mark=diamond*\\%
}

\DeclareRobustCommand\legendbox[1]{(\textcolor{#1}{#1}~\begin{tikzpicture}[x=0.2cm, y=0.2cm] \draw [color=black, fill=#1!20] (0,0) -- (0,1) -- (0.6,1) -- (0.6,0) -- (0, 0); \end{tikzpicture})}

\pgfplotsset{
SmallBarPlot/.style={
    font=\scriptsize,
    ybar,
    width=\linewidth,
    ymin=0,
    xtick=data,
    xticklabel style={text width=1.5cm, align=center}
},
SmallBarPlotHorizontal/.style={
    font=\scriptsize,
    xbar,
    width=\linewidth,
    ytick=data,
    yticklabel style={text width=1.5cm, align=center}
},
BlueBars/.style={
    fill=blue!20, bar width=0.25
},
RedBars/.style={
    fill=red!20, bar width=0.25
},
BlueBars2/.style={
    fill=blue!20, bar width=0.3
},
RedBars2/.style={
    fill=red!20, bar width=0.2
},
GreenBars/.style={
    fill=green!20, bar width=0.2
},
OrangeBars/.style={
    fill=orange!20, bar width=0.2
}
}

\usepackage[textsize=tiny, textwidth=3cm]{todonotes}

\usepackage[hidelinks, breaklinks]{hyperref}

\usepackage{booktabs}
\newcommand{\calE}[0]{\mathcal{E}}
\newcommand{\calF}[0]{\mathcal{F}}

\newcommand{\calW}[0]{\mathcal{W}}
\newcommand{\calP}[0]{\mathcal{P}}

\SetKwInput{KwInput}{Input}                
\SetKwInput{KwOutput}{Output}              

\usepackage{diagbox} 

\usepackage[normalem]{ulem}

\newcolumntype{M}[1]{>{\centering\arraybackslash}m{#1}}

\usepackage{hyperref}

\AtBeginDocument{%
  \providecommand\BibTeX{{%
    \normalfont B\kern-0.5em{\scshape i\kern-0.25em b}\kern-0.8em\TeX}}}

\begin{document}

\begin{frontmatter}              


\title{Adversarial Reconnaissance\\Mitigation and Modeling}
\runningtitle{Adversarial Reconnaissance Mitigation}

\author[A]{\fnms{Shanto} \snm{Roy}%
\thanks{Corresponding Author: Shanto Roy, University of Houston; E-mail:
shantoroy@ieee.org}},
\author[B]{\fnms{Nazia} \snm{Sharmin}},
\author[B]{\fnms{Mohammad Sujan} \snm{Miah}},
\author[C]{\fnms{Jaime C} \snm{Acosta}},
\author[B]{\fnms{Christopher} \snm{Kiekintveld}},
\and 
\author[D]{\fnms{Aron} \snm{Laszka}}

\runningauthor{first\_author et al.}
\address[A]{University of Houston, TX, USA}
\address[B]{University of Texas at El Paso, TX, USA}
\address[C]{DEVCOM Army Research Laboratory, MD, USA}
\address[D]{Pennsylvania State University, PA, USA}

\begin{abstract}
Adversarial reconnaissance is a crucial step in sophisticated cyber-attacks as it enables threat actors to find the weakest points of otherwise well-defended systems.
To thwart reconnaissance, defenders can employ cyber deception techniques, such as deploying honeypots.
In recent years, researchers have made great strides in developing game-theoretic models to find optimal deception strategies. 
However, most of these game-theoretic models build on relatively simple models of adversarial reconnaissance---even though reconnaissance should be a focus point as the very purpose of deception is to thwart reconnaissance.
In this paper, we first discuss effective cyber reconnaissance mitigation techniques including deception strategies and beyond. Then we provide a review of the literature on deception games from the perspective of modeling adversarial reconnaissance, 
highlighting key aspects of  reconnaissance that have not been adequately captured in prior work.
We then describe a probability-theory based model of the adversaries' belief formation and illustrate using numerical examples that this model can capture key aspects of adversarial reconnaissance.
We believe that our review and belief model can serve as a stepping stone for developing more realistic and practical deception games.
\end{abstract}


\end{frontmatter}

\thispagestyle{empty}
\pagestyle{headings}


\section{Introduction}\label{sec:intro}

Reconnaissance is necessary for adversaries to understand network topology and target, identify vulnerabilities, and plan attack vectors accordingly before exploitation. After compromising an internal asset, an adversary can perceive network information (network hosts, operating system) using active or passive reconnaissance. 


Defenders can use static or dynamic deception (e.g., honeypot placement or OS obfuscation) to prevent effective adversarial reconnaissance leading to misinformation or uncertainty about host or network configurations. 
Deception strategies including perturbation, moving target defense, obfuscation, mixing, honey-x, and attacker engagement can deceive and delay reconnaissance~\cite{pawlick2019game}.
However, very few of them model the reconnaissance as a belief update process that must integrate multiple (potentially conflicting) observations over time, which leads to decision and game theory models that are not able to fully capture the detailed effects of deception on attacker planning. We study the following research questions:

\begin{itemize}
    \item[\textbf{Q1.}] \textbf{Adversarial Reconnaissance Mitigation:} How to effectively mitigate adversarial reconnaissance?

    \item[\textbf{Q2.}] \textbf{Adversarial Reconnaissance Models in Literature:} How is adversarial reconnaissance defined and modeled in the existing literature on game theory for cyber deception?
    \item [\textbf{Q3.}] \textbf{Adversarial Belief Formation:} How to more accurately model incomplete and evolving attacker beliefs based on realistic observations, and how could this lead to more effective deception strategies? 

\end{itemize}


We answer the first question by discussing research evidence, examples, and practices (Section~\ref{sec: recon-mitigate}) both from academia and industry. In this section, we discuss \emph{Cyber Deception} and \emph{Moving Target Defense} that are being adapted both in academic research and industrial implementation. The combination of deception and movement can potentially mitigate the reconnaissance process and eventually can delay adversarial approaches. We also discuss \emph{security awareness} and \emph{security best practices} that mitigates or prevents reconnaissance techniques to be successful.

We answer the second question by surveying representative work from the existing literature from the perspective of cyber deception to identify the typical approaches for defining and modeling adversarial reconnaissance (Section~\ref{sec:recon}).
Most deception games model reconnaissance as a simple procedure that always reveals certain network information, and optimizes deception strategies accordingly.
These game-theoretic models also calculate the attacker's action based on observations (e.g., scan results) without considering how long an attacker waits or how beliefs evolve over time in a deception-enabled network.

For the third question we develop an initial Bayesian belief update model for an attacker based on a knowledge base and a number of observations (Section~\ref{sec:passive_recon_model}). In this model, (i) we consider reconnaissance about the host configurations (operating system, application, or service), (ii) we model the attacker's observations and knowledge base, and (iii) we use the Bayesian belief model to update the attacker's beliefs about configurations. We propose that such a model can be used to find different deceptions strategies based on the details of the attacker's observations and current beliefs. 
Finally, we provide concluding remarks and future perspectives in Section~\ref{sec:conclusion}.



\section{Adversarial Reconnaissance}\label{sec:background}
We begin by surveying related work on adversarial reconnaissance methods and models in the context of cyber deception.

\subsection{Adversarial Reconnaissance Techniques} 
Adversaries conduct reconnaissance to identify attack vectors in targeted systems.
Reconnaissance techniques fall primarily into two categories: active and passive. Active methods require interaction between the attacker and the environment, e.g., sending messages and interpreting responses to scan a target network or host.  In contrast, passive methods do not require interaction; the adversaries quietly observe network data (e.g., captured network packets), analyze these data, and use it to form beliefs about possible network configurations. 
For example, fingerprinting is done by observing traffic data on a network without any active modification to detect active nodes, virtual hosts, opened ports, address ranges, network addresses, functional machines, operating systems, running services, etc.~\cite{roy2022survey}.

There are many tools used for performing reconnaissance within the network. While scanning the TCP/IP stack or OS fingerprinting, the widely used tools include nmap, xprobe2, and p0f. Particular tools (e.g., nikto, dirbuster) can be used to perform reconnaissance in the application layer~\cite{fraunholz2018catch}. 
Kiekintveld et al. mention Nessus and OpenVAS for performing port scanning to find installed software versions and similar information~\cite{kiekintveld2015game}. They also mentioned tools like MulVAL and NetSPA which can be utilized to create an attack graph (though many of these tools require administrative access and are commonly used by defenders). Durkota et al. also considers attack decisions based on the scores provided by Common Vulnerability Scoring System~\cite{mell2006common} and National Vulnerability Database~\cite{bacic2006mulval}, in combination with an attack graph~\cite{durkota2015optimal}. Finally, Roy et al. presented a taxonomy of adversarial reconnaissance techniques alongside other taxonomies including reconnaissance target information and phases during an attack chain~\cite{roy2022survey}.

\subsection{Adversarial Reconnaissance Tools}
There are a number of information gathering tools to perform reconnaissance. Dar et al. explored a range of reconnaissance methods that attackers employ to learn more about the target~\cite{dar2018silent}. They assessed which recon tactic, while concealing the identity of the enemy, gets the most data about the target. They experimented with a variety of operating systems and looked into the tools DNSENUM, NMAP, ZENMAP, DNSstuff, and MxToolbox. They came to the conclusion that if an attacker wanted to remain anonymous, DNSENUM, NMap, and ZENMAP performed well in active reconnaissance, while DNSstuff and MxToolbox performed well in passive reconnaissance.

Research on stealth port scanning techniques was done by Claypool~\cite{claypool2002stealth}. He concentrated on techniques like NMap and Hping2, as well as half-open, Xmas tree, UDP, Null, Fragmentation, Decoying, and Spoofing, which are known for stealthy port scanning. The investigations offer a qualitative examination of these methods and equipment. Another study by Tundis et al.\ also focused on vulnerability tools~\cite{tundis2018review}. However, the work is based on qualitative analysis only. Most above-mentioned works analyze the advantages and limitations of the tools in different scenarios and assess the performance of these tools within enterprise network.

\section{Reconnaissance Mitigation and Prevention
}\label{sec: recon-mitigate}
 

Defensive measures against the reconnaissance techniques differ from one another based on the approaches, target people or platform, and the tools adversaries use. Defenders can attempt to mitigate a few reconnaissance techniques while in other cases mitigation and prevention measures are not effective. For example, it is quite difficult to prevent internet footprinting as organizations keep exposing themselves through marketing. Besides, information exposure largely depends on the people working within the organization and they are equally responsible to keep confidential information safe and out of the reach of adversaries. Social engineering is a psychological game between two parties and one is the ultimate winner. Social engineering has no technical defensive measures and only sufficient training and security awareness can mitigate it at large. Network scanning techniques are a bit cautious from the adversary's perspective. Therefore, defenders can utilize enforced firewall rules, intrusion detection and prevention measures, cyber deception, and moving target defense technologies to mitigate and prevent the effectiveness of scanning techniques. Local discoveries have limited
preventive measures as 
discovery involves obtaining local information in a compromised host through overwriting system configurations and basic command-line tools and techniques. 
Regardless of the different techniques utilized by adversaries, organizations take certain measures to defend their resources. Defense involves physical precautions, use of firewalls and intrusion detection systems (e.g., host or network-based IDS), enforcement of security training to employees, secure network administration, zoning, and other best practices. Apart from the traditional security measures, both academia and the industry is adapting to modern defensive techniques such as deception and moving target defenses (MTD) intending to deceive the attacker and make the search or operational space more complex.

\begin{figure*}[!ht]
    \centering
    \includegraphics[width=\textwidth]{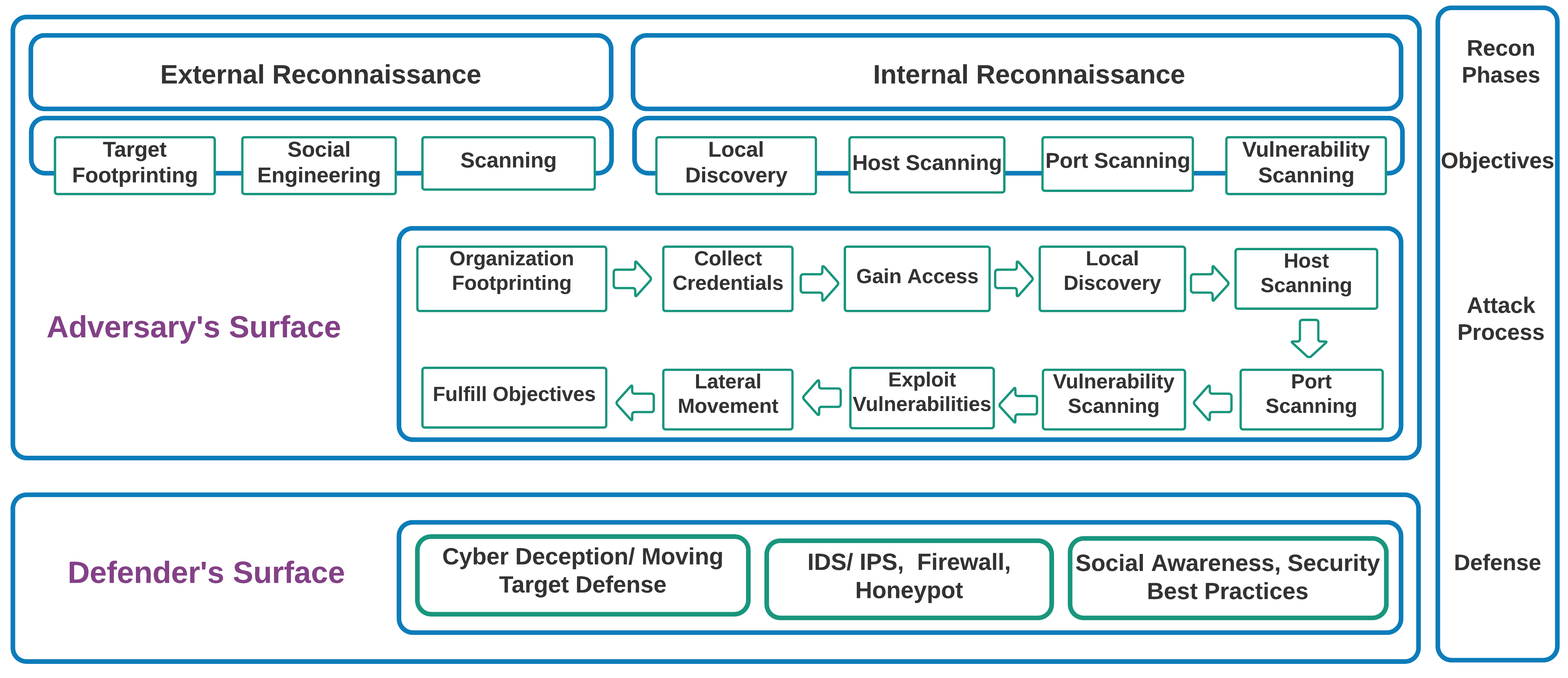}
    \caption{Adversary's Surface vs. Defender's Surface}
    \label{fig:attack_surface}
\end{figure*}

Figure~\ref{fig:attack_surface} shows to the attack process from the adversary's perspective and the defensive measures from the defender's perspective. The figure depicts how adversaries approach based on the external and internal reconnaissance and in contrast, what defensive measures organizations can utilize.

\subsection{Reconnaissance Detection}

In order to mitigate and limit adversarial reconnaissance, defenders are required to detect the actions, behaviors, and patterns of these techniques. \emph{Target footprinting} techniques can rarely be detected though as it is passive and other people (e.g., employees, network administrator, business competitor, job seekers) can look for publicly available organization information as well. The only way to mitigate target footprinting is understanding the privacy level (confidential or public) of data before publishing or posting in a data source (websites, documents, magazines, etc.).

\emph{Social engineering} techniques are combination of different confidence tricks that are difficult for the defender (a person or a group of people) to identify against a skilled attacker. However, effective understanding of the context in terms of the requester identity (authority level, credibility, previous existence, verifiable, etc.) and data (sensitivity, necessity to provide, public/private domain, etc.) can help to identify such attacks at large~\cite{mouton2015social}. Phishing is one of most common form of social engineering and according to Nicholson et al.,\ defender person can identify a phishing email better when they have more attention to the sender's details~\cite{nicholson2017can}. However, people who make quick decisions are prone to fall victim to the phishing emails. Sawa et al.\ utilizes natural language processing to identify typical suspicious questions and commands used by adversaries~\cite{sawa2016detection}. Ahmed et al.\ differentiates phishing URLs from regular ones by developing an algorithm that checks for URL length, number of ``//'' appear in the URL, if contains ``@'' or not, and if there is ``-'' in prefix or suffix~\cite{ahmed2016real}. Apart from these detection and mitigation techniques, mobile users can take typical measures such as content filtering, blacklisting, and whitelisting of web addresses to tackle phishing, smishing, and vishing~\cite{shahriar2015mobile}.

\emph{Network scanning} techniques are performed frequently from an external network and it is more robust within the internal network once an internal host is compromised. System and network administrators are required to protect the organization network from these scans because this is how adversaries get to know about active hosts, ports, and vulnerabilities. However, recognizing the scans performed by adversaries are typically a strenuous task for administrators and adversaries have been innovating new strategies to avoid detection.
Bhuyan et al.\ proposed two approaches towards detecting severe port scans such as \emph{single-source} and \emph{distributed} port detection~\cite{bhuyan2011surveying}. The single source scan can be detected at \emph{flow-level} or \emph{packet-level}. Authors have also categorized single source scan detection as \emph{algorithmic approaches} that use hypothesis testing and probabilistic models to detect the scans (e.g., detection using network packet-based activity graph~\cite{staniford1996grids}, anomaly-based detection~\cite{kim2004detecting,bhuyan2017network}), \emph{threshold-based approaches} that use a fixed-size time window to detect how many unique IP addresses have already been contacted by the host (e.g., checking number of SYN packets over a time~\cite{roesch1999snort}, number of destinations contacted by a single source~\cite{paxson1999bro}, time-based flow size distribution~\cite{zhang2009novel}), \emph{soft-computing approaches} that emphasize flexible information processing to trace the scan origin (e.g., detection based on learning from past histories of open ports~\cite{chen2009novel}, \cite{liu2008network}), \emph{rule-based approaches} that utilize stored rules in the knowledge-base to detect and prevent typical scans (e.g., abnormal traffic control framework~\cite{kim2008slow}), and \emph{visual approaches} that leverage visualization to detect a particular network event like scanning (e.g., visualizing patterns of scan generated by Nmap, Nessus, etc.~\cite{conti2004passive}). In contrast, authors claimed that distributed scans can be tracked at \emph{packet-level} and \emph{alert-level}. Different algorithmic, soft-computing, and visual approaches in a distributed mode, and clustering-based methods can be used to detect distributed scans~\cite{robertson2003surveillance,staniford2002practical,gates2006co}.

Kim et al.\ proposed a scanning detection mechanism based on fuzzy logic and a stepwise policy (decreasing network bandwidth $\rightarrow$ discarding packets)~\cite{kim2008slow}. Their framework had an intrusion detection module consisted of a ``packet analysis'' and an ``intrusion analysis'' sub-modules that analyze the network packets using fuzzy logic and inspects intrusion respectively. The intrusion prevention module included a ``packet filtering'' sub-module that blocks abnormal packets and a ``queue assignment'' sub-module that narrows the bandwidth if a series of packets are found suspicious.  
Scanning using botnets make it more convenient for adversaries to hide the true identity and the origin of an attack. Botnets can recruit vulnerable hosts to spread and launch scanning attacks. Botnets are robust in a large scale attack and adversaries can execute commands through a C\&C channel. As these bots are used to perform several scanning algorithms (e.g., uniform, non-uniform, localized), it is necessary to detect botnet commands and compromised host. Gu et al.\ utilized statistical analysis to detect spatial-temporal correlation and similarity with the previous real-world botnets so that they can capture the engagement by analyzing the network traffic~\cite{gu2008botsniffer}.

Vulnerability scan detection is typically strenuous in a network if it is limited to passive banner grabbing. However, active scanning can be detected in a server if adversaries leave logs behind. Seyyar et al.\ proposed to detect potential web vulnerability scan by looking into the access log files of different web servers (e.g., Apache, IIS)~\cite{seyyar2018detection}. The rule-based detection works by examining the past recorded logs using existing vulnerability scanners and analyzing suspicious activities. 

\emph{Local discovery} techniques are performed when adversaries already have access to the system and if privileges are escalated, there is less chance to detect and mitigate the reconnaissance. However, using host-based intrusion detection techniques and security softwares (antivirus or firewall) can be efficient against inexperienced attackers.

\subsection{Cyber Deception and Moving Target Defense}
The traditional mechanism against reconnaissance lies in customized network design, firewalls, network address translations (NAT), DMZ, subnetting, etc. However, traditional network management does not guarantee security due to misconfigurations, constant scenario changes, and other human errors. Therefore, next-generation concepts such as \emph{Cyber Deception} and \emph{Moving Target Defense} (MTD) have evolved in the network security scenario; bringing more features to mitigate adversarial scanning and discoveries.

Cyber deception and MTD adapts to the adversary's behavior and manipulate their behavior by creating static or dynamic deception and diversion. Wang et al.\ differentiate cyber deception from MTD 
based on static pre-configuration and misdirection by providing misleading information while MTD focuses on dynamic approaches by increasing complexity and diversity on the fly \cite{wang2018cyber}. Crouse et al. elaborate the differentiation by providing examples for each approach: deception includes network masking, decoying, mimicking, and inventing (honeypots); on the other hand, MTD includes dynamic network shuffle, configuration adaptation, and evolution, and Address Space Layout Randomization (ASLR) \cite{crouse2015probabilistic}. ASLR involves random rearrangement of address spaces positions of a process in the memory to prevent successful code or script execution. 
Crouse et al.\ also claimed that, reconnaissance defense using deception or movement costs the attacker more than the defender making the defenses potentially successful. 

Han et al.\ have surveyed and listed several deception techniques that are able to mitigate, detect or prevent adversarial reconnaissance~\cite{han2018deception}. The list includes deceptive topology, OS obfuscation, traffic forging, decoy services, etc. as network-layer deception; decoy computation, multi-layer deception, fake honeypots, honey-permission, etc. as system-level deception; software decoys, honey-patches, shadow honeypots, honey-URL, honey-accounts etc. as application-level deception; and honey-passwords, honey-encryption, honey-accounts, honey-files, honey-profiles and webpages etc. as data-level deception.
The earlier stage of cyber deception grew mostly from the tactics of military deceptions. Trassare et al.\ proposed a proof-of-concept network to deceive the UDP-based ``traceroute'' query by shunting the incoming packet to a deceptive topology~\cite{trassare2013technique}.
Deploying honeypots in the network can create misdirection and eventually can detect intrusion. However, Wang et al.\ claim that botnet attacks are very common nowadays and adversaries are capable of identifying the honeypot nodes in both centralized and peer-to-peer (P2P) botnets~\cite{wang2010honeypot}. Therefore, researchers have come with different solutions to address this problem. Challoo et al.\ experimented to detect adversarial botnets using honeypots by differentiating between the infected honeypot and the uninfected one~\cite{challoo2011detection}. On the other hand, Al-Hakbani et al.\ utilize peer lists and vulnerabilities information to prevent the Advanced Two-Stage Reconnaissance Worm (ASTRW) by remotely controlling the infected hosts~\cite{al2015avoiding}.

Not only the academia but also the global industry is adapting to the deception technologies. According to the Global cyber deception report of 2019 \cite{GlobalCy21:online}, a few of the key players in the market who already have developed security systems adapting to deception technologies are Rapid7\footnote{\url{https://www.rapid7.com/products/insightidr/features/deception-technology/}},
LogRhythm\footnote{\url{https://logrhythm.com/}},
Trapx Security (DeceptionGrid\footnote{\url{https://trapx.com/product/}}),
Attivo Networks\footnote{\url{https://attivonetworks.com/deception-based-threat-detection/}},
Illusive Networks\footnote{\url{https://www.illusivenetworks.com/}},
Smokescreen Technologies\footnote{\url{https://www.smokescreen.io/}},
Acalvio Technologies\footnote{\url{https://www.acalvio.com/}},
etc.

\subsubsection{Software-Defined Networking in Deception and MTD}
Software-Defined Networking (SDN) has been utilized well in both deception and movement strategies.
Shimanaka et al.\ proposed a system that creates an identically configured network for deceiving adversarial reconnaissance~\cite{shimanaka2019cyber}. The system was designed intending to minimize further compromise once intrusion is detected by redirecting all traffic from the operational network (O-net) to deception network (D-net).
SDN can deceive adversaries by simulating virtual networks and can defend from potential insider reconnaissance as well. Achleitner et al.\ developed  a reconnaissance deception system (RDS) that delays the adversarial scanning and invalidates their already collected information over time~\cite{achleitner2016cyber}~\cite{achleitner2017deceiving}. The simulation can mitigate network reconnaissance techniques such as vulnerable host scanning by APTs. With the evolve of network deception techniques utilizing SDN, Robertson et al.\ developed \emph{CINDAM} that creates modified per-host view to make the network topology more unpredictable~\cite{robertson2015cindam}. CINDAM was cost-effective and achieved this deception without affecting any visible performance issue in the internal network. Later, Chiang et al.\ developed an adaptive cyber deception system entitled \emph{ACyDS} that changes the network view of each host dynamically~\cite{chiang2016acyds}. To build the network deception more adaptive, Sugrim et al.\ proposes to combine deception and MTD; intending to present a \emph{dynamic virtual attack surface}~\cite{sugrim2018measuring}. This virtual surface features isolated view for each host, dynamic change of view, and large search space (subnets with large number of hosts). 

Wang et al.\ developed a method using SDN based MTD, Sniffer Reflector to prevent adversarial network reconnaissance. Sniffer Reflector simply creates an obfuscated view for adversaries to refrain them from obtaining the original result~\cite{wang2016moving}. Shi et al.\ developed another SDN-based MTD approach called CHAOS that features IP, ports, and fingerprint obfuscation to present the network environment more unpredictable~\cite{shi2017chaos}. As fingerprinting is severe within the internal network, Zhao et al. have developed a signal game utilizing the SDN-based fingerprint hopping (FPH) method to counter adversarial fingerprinting techniques \cite{zhao2017sdn}. To prevent scanning techniques, Wang et al. have proposed random domain name and address mutation (RDAM) that increases the address search space by dynamically changing or remapping the IP addresses in a network \cite{wang2017random}.

\subsubsection{Techniques and Components of Deception and MTD Deployment}
Network virtualization and network function virtualization can easily be utilized to create dynamic views from each host, topology deception and creating large scaling search spaces using low-cost virtual machines. Both Deception and MTD targets to mislead adversaries by providing false information and direction within the network. Deception and MTD introduce different techniques and tools so far that are utilized by altering static and dynamic configurations in the network.

\paragraph{Deception and MTD Techniques}
Several deception and MTD techniques have been utilized to-date and researchers have been introducing new techniques for efficient deception/MTD-based defense. After SDN has been introduced, deploying these techniques have become easier. 
Deception techniques can be applied in different layers such as network layer, data layer, application layer and phases of the network system. 
The following deception techniques are appropriate for insider adversaries who have placed themselves in the network (on one or more hosts), using techniques such as social engineering, exploiting zero-day vulnerabilities, via drive-by downloads or by manual infection, etc.

\begin{itemize}
    \item\textbf{Dynamic address translation (DAT):} 
    To make the overall address space of a network appear larger on-the-fly packet header rewriting can be used to hide the real host addresses which make the network search space big. DAT can be implemented by both hardware and software through the use of page tables, segment tables, region tables.  Additionally, allows different address spaces to share the same program or other data that is for read-only. The translation of the real underlying network’s address space into a significantly larger  address space increases the search space for adversarial scanners and invalidate their gathered information
\cite{achleitner2017deceiving,achleitner2016cyber}.

\item\textbf{Route alteration:}  By simulating virtual paths, multiple hops from a source to a destination host can be spanned \cite{achleitner2016cyber}. Which  can alter the topology of different network views so that a scanner will be unable to correctly conclude about the real network topology~\cite{borders2007openfire}.

\item\textbf{Address space randomization:}
The IP address space of the target network including both the real servers and the decoys can be shuffled~\cite{sun2016desir}. Although through reconnaissance, the attacker can create a blacklist of decoy IP addresses, this blacklist will not be valid after the next round of IP randomization, and the attacker has to start over the reconnaissance process.

\item\textbf{Forwarding ARP requests to deception server:}
ARP packet is an important part of the deception system. ARP requests are usually flooded into the network to discover hosts and match IP to MAC addresses. By including a deception server which can handle all ARP requests forwarded to it and sends the appropriate response packets which can ensure that hosts which are not supposed to be discovered stay hidden \cite{rowe2004model,achleitner2016cyber}.

\item\textbf{Resource obfuscation:}
Host-based (OS) obfuscation, ports, address space, and obfuscation based on decoys can mitigate internal recon by making the network more unpredictable~\cite{shi2017chaos}. Resource obfuscation through migration and dynamic mapping provide a complex large search space for adversaries.
\end{itemize}

\paragraph{Deception and MTD Components}
Several tools and components are considered useful in terms of delaying and preventing adversarial reconnaissance. These components effectively complement other well-known intrusion detection and prevention techniques as well. Moreover, it improves on the more conventional approaches that the security community has studied and implemented during the past 35 years\cite{wang2018cyber}. Here we discuss a few common tools used for cyber deception and MTD:

\subparagraph{Honeypots} Honeypots are 
host machine frameworks intended to be attacked so that it helps network administrator to analyze the attack pattern without having any damage to other real network nodes. It can be used in different aspects of security such as prevention, detection, and information gathering. It is a tool that is flexible in nature and with applications in many areas of the network. There are two particular honeypot types:
  
\begin{itemize}
\item \textbf{High-interaction honeypot:} It simulates all aspects of an operating system. Via High Interaction honey pot attackers can gain full access to the system
\item \textbf{Low-interaction honeypots:} It can simulate only some parts, such as the network stack.Via low-interaction honeypots adversaries can simulate only services that cannot be exploited to get complete access to the system \cite{shulman2015compromised}.
\end{itemize}

The key advantage of using honeypots is to mislead adversaries to decoy systems and to detect an intrusion in a system or network. Besides, using honeypots the size of a search space and sparsity of real vulnerabilities can be increased. Apart from these, a deep examination of adversaries during and after exploitation of a honeypot can be analyzed~\cite{mokube2007honeypots} for considering further security measures. However, honeypots can be used as a launching platform to attack other hosts on different network~\cite{mokube2007honeypots}.



\subparagraph{Honeytools}
The prefix ``honey'' has been utilized to allude to a wide scope of methods that fuse the demonstration of deceiving in them. The fundamental thought behind the utilization of the prefix word ``honey'' in these methods is that they have to allure adversaries to interact with them.

    \begin{itemize}
    \item \textbf{Honeyfiles:} Honey files are bogus files and folders that make real files and folders indistinguishable from fake files and folders. After the attacker intrudes inside the network and compromises a machine, he can escalate his privilege into the next stage of an attack, where local emails, files, and folders are explored. Honey files (HFs) can play a deception role in this phase. Honeyfiles can be perfectly believable decoys or the modification of authenticating files to include some alteration in the function~\cite{provos2004virtual}.

    \item \textbf{Honeytokens:} This is similar to honey flies. These are fake data that attracts attacker to interact with the data, for example, DNS honey token is a complementary technique to honeypots. Whenever adversaries use “brute force” for common subdomains or attempt a zone transfer, DNS servers will try to identify interesting resources (e.g., sub-domains, servers) as part of their reconnaissance process. By creating a small number of fake DNS records on the authoritative DNS servers of the organization and by configuring them to initiate an alert when these specific records are requested; defenders can receive an early warning of DNS-related information-gathering attempts against their infrastructure. Another example, fake credit card numbers can be inserted into the repository within a network. IDS’s can be configured to watch the network if these are accessed then the notification of system compromised will be sent to defender~\cite{mokube2007honeypots}.
    
    \item \textbf{HoneyGen:} Honeygen is usually used to produce honeytokens automatically. HoneyGen produces honeytokens that are similar to the real data by deriving the characteristics and properties of real data. These honeytokens can be used to detect intrusions, data misuse and data leakage by adversaries. Honey tokens can be implemented among real tokens in the database and can be monitored to detect any activity associated with them~\cite{bercovitch2011honeygen}.
    
    \item\textbf{Honeywords:} Honeywords or honey-passwords are typical fake passwords, utilized to confuse adversaries in finding the difference between real and fake passwords when they crack a stolen hashed password file. Honey passwords are having multiple possible passwords for each account, only one of which is genuine~\cite{mokube2007honeypots}.
    
    \item\textbf{Honeyd:} A framework for virtual honeypots to simulate virtual computer systems at the network level. Honeyd can simulate the networking stack of various operating systems and able to provide arbitrary services for any number of virtual systems~\cite{provos2003honeyd}.

\end{itemize}

The usage of the above-mentioned tools can significantly increase the possibility of detecting potential attacks at different reconnaissance phases. 
Combining these tools and techniques make reconnaissance, penetration and lateral movements further difficult for adversaries. However, the effectiveness of such measures against insiders is not measured yet as it is likely they know the use of these tools.




\subsubsection{Game Theory for Deception and MTD}
A game model can play a vital role to increase the uncertainty of adversarial reconnaissance, and also can give different efficient strategies to the defender to implement deception effectively. Deception is an important tactic against adversary reconnaissance, and there are different approaches that apply game-theoretic analysis to cyber deception. The network administrator can introduce deception into response networks scans, such as obscuring certain system characteristics, the game model helps to find defender strategy to implement deception more effectively. Game theory can analyze and develop a solution for decision-making problems among a group of players. It gives quantitative reasoning where the players are either uninformed or unknown about the strategies, moves, and intent of opponents. Therefore, the game model can provide optimal strategies for both defender and attackers to predict the outcome of the deployment of various deception strategies. The Cyber Deception Game represents the Stackelberg game between the defender (e.g., network administrator) and an adversary where the defender moves first and chooses how the systems should respond to scan queries~\cite{schlenker2018deceiving}. A defender can introduce deception by introducing  observed configuration which will hide the true configuration of the network and reduce the attacker's success in reconnaissance phase.

The defender can  protect the true system, from possible exploits and intrusions.\
Each system has certain attributes, e.g., an operating system, a services host, etc. However, when an attacker will view the system it will see the observed configuration which is how a system administrator ought to react to scan requests from an attacker endeavoring to penetrate the system.   Mixed integer linear programming is proved efficient to find strategies of defender against the powerful attacker.  Additionally, greedy algorithm can be used which quickly finds good defender strategies and performs well empirically \cite{schlenker2018deceiving}. Honeypot selection game can leverage the network configuration for a target, by introducing deception game model, the defender can use honeypots in the network in a way so that the attacker will choose to attack honeypots instead of real machines and reduce the chance of correct information gathering in reconnaissance phase.  Probing actions can be used by the attackers, where attackers tried to distinguish which one honeypots from real machines before launching an attack. The probes were noisy, that's why attackers used imperfect information in these models. Using linear programming models for finding the best strategies for the defender to maximize deception strategies can maximize defenders gain in deceiving adversaries in reconnaissance phase Radek Píbil et al.~\cite{pibil2012game}. There are also some works that focused on how to effectively use honeypots (fake systems) as part of a network deception \cite{carroll2011game,wagener2009self,pibil2012game}. 

There has also been a different work on security games that focuses on the role of deception in general to manipulate the beliefs of an attacker~\cite{yin2013optimal,an2011refinement,horak2017manipulating,thakoor2019general}. Finally, to understand the effectiveness of cyber deception and how different research contributed to deploy cyber deception and gain effectiveness in deception game theory can play a predominant role in deceiving attackers at the reconnaissance phases~\cite{ferguson2019game}.

\subsection{Security Awareness and Best Practices}
Security awareness requires both ``knowing'' and ``doing'' the right security actions and measures~\cite{Cybersec84:online}. Most of the security attack is caused by human error (e.g., misconfigured server or database, vulnerable source codes and scripts) and according to the IBM threat landscape of 2018, the human error factor caused 
$43\%$ of publicly disclosed misconfiguration incidents~\cite{IBMXForc60:online}. For example, developers can misconfigure a configuration file and leave security holes to be exploited. This is why adversaries can be benefited a lot by performing social engineering techniques. Simple passwords can easily be cracked using brute-force or dictionary attacks. Organizations should provide sufficient security awareness training toward the users and here we discuss a few security best practices:


\begin{itemize}
    \item \textbf{Use Strong Password and Passphrase:}
    Many people use weak passwords (e.g., shorter length, words in a dictionary, birth year, mobile number) which is easy to break using Bruteforce or Dictionary attack. Users should use a minimum 8-character length password using numeric values, special symbols (\$, @, \&, \%, etc.), capital and small case letters. Users can also use passphrases that are easier to remember and longer as well. Passphrases are the combination of multiple words separated by ``space'' character (a sentence or a phrase). Strong passwords or passphrases can prevent adversaries to guess or to break potential user credentials.
    
    \item \textbf{Two-Factor Authentication or Two-Step Verification:}
    Users need to maintain two-factor authentication~\cite{jin2004biohashing} or two-step verification~\cite{van2009twostep} while accessing resources or personal accounts. Two-factor authentication requires something users remember (password), something they are (biometric scan, e.g., fingerprint/retina) and something they have (another trusted device, e.g., mobile/computer)~\cite{Twofacto9:online}. On the other hand, two-step verification involves users providing two sequential tokens- first the password, then a One-Time Password (OTP) sent over SMS or email address. Using these security mechanisms prevent unauthorized access even after adversaries extract valuable user credentials.
    
    \item \textbf{Limit BYOD:}
    BYOD stands for Bring Your Own Device. Allowing BYOD in an organization creates opportunities for malicious insiders who can connect an infected device to the internal network. Users can also connect infected pendrives or hard-drives unknowingly in the network. Therefore, organizations can counter the social engineering techniques, \emph{Baiting} and \emph{Reconnaissance by Insider} in particular by limiting the use of personal devices within the network. 
    
    \item \textbf{Utilize IDS and IPS:}
    Organizations should introduce intrusion detection and prevention systems to identify typical penetration in the network. Host-based IDS and Network-based IDS are the most common detection systems. HIDS can mitigate the risk of local discovery techniques in a compromised host by identifying the presence of an attacker. On the other hand, NIDS can mitigate and slow down the adversary's approach while performing network scanning techniques. Honeypots can be utilized as well by deploying it within the network with particular configurations. 
    
    \item \textbf{Enforce Firewall Rules:}
    Organizations need to enforce certain firewall rules that block particular scanning packets (TCP/UDP/ICMP) and mitigate the network-based reconnaissance techniques. Organizations should use both hardware and software-based firewalls as a security measure. Hardware firewalls are typically used to protect a network and are deployed at the entrance of the network (e.g., a gateway). Software firewalls are installed in a device protecting it from other internal or external nodes. Typical rules can avoid particular ICMP, TCP, and UDP network scan requests.
    
    \item \textbf{Isolate Zones:}
    Dividing the whole network into separate zones can block and slow down adversaries to have a whole picture of the network. Isolated zones are useful in case BYOD is allowed in an organization and can easily be adapted to cyber deception and moving target defense. Therefore, even a device is compromised adversaries cannot have the whole picture of the entire organizational network.
    
    \item \textbf{Resource Access Control:}
    Organizations should maintain proper access controls to resources in terms of both physical and network-based access. Proper access control can slow down the adversaries while performing internal scanning and discoveries.
    
    \item \textbf{Vulnerability Inspection:}
    Organizations are often (e.g., each month) required to inspect vulnerabilities within the system to identify any misconfiguration or security holes. Performing SWOT (Strength-Weakness-Opportunity-Threat) analysis on regular-basis can mitigate the effectiveness of reconnaissance at different levels and phases.
    
    \item \textbf{Update Patches:}
    Regular update of patches can mitigate the effect of vulnerability scanning. Organizations should keep all the security measures (e.g., security softwares/antivirus) up-to-date to minimize vulnerability exposure and discovery techniques in the local machine.
    
    \item \textbf{File Encryption:}
    As adversaries are capable of accessing files, users should encrypt files containing confidential information in order to keep information secure and safe. Encrypting files and directories can mitigate local file discovery techniques.
    
    
\end{itemize}

\section{Reconnaissance Models in Deception Games}\label{sec:recon}

Deception game models are useful tools to investigate the effectiveness of cyber deception in the reconnaissance phase and find the optimal strategy for deploying cyber decoys~\cite{ferguson2019game}. 
We first provide on overview (using a representative sample of the literature) of how existing work on deception games models adversarial reconnaissance (Section~\ref{subsec:recon_lit}). Then we assess the limitations of the models of adversary beliefs in existing work (Section~\ref{subsec:lit_limitation}). Finally, we discuss the most closely related works focusing on how better adversary models could result in better deception strategies (Section~\ref{subsec:non-game-theory-models}).

\subsection{Reconnaissance in Deception Games}\label{subsec:recon_lit}

\begin{figure}[!ht]
    \centering
    \includegraphics[width=.65\textwidth]{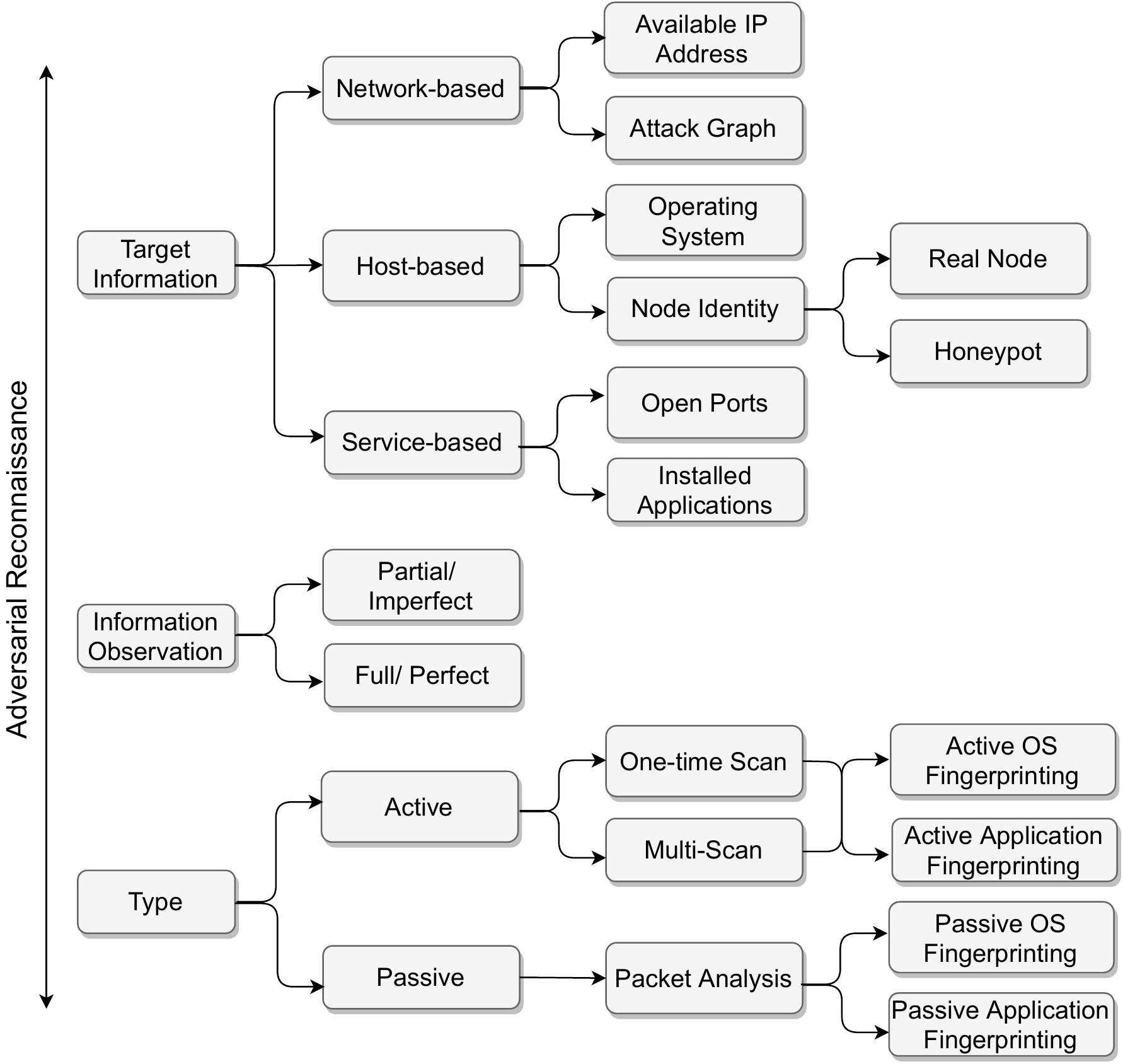}
    \caption{High-level taxonomy of adversarial reconnaissance.
    }
    \label{fig:recon_taxonomy}
\end{figure}

We give a high-level overview of adversarial reconnaissance (Figure~\ref{fig:recon_taxonomy}) and then discuss several categories of models found in the deception literature.

\subsubsection{Reconnaissance Models}
Poschinger et al. discussed reconnaissance as finding IP addresses (utilizing host or port scanning) using Nmap~\cite{poschinger2020openmtd}. 
Many game-theoretic models are related to attack graphs, honeypot placement, or masking node identity~\cite{anwar2020game,milani2020harnessing} that consider the host as the primary target. 
Some game models consider how efficient the reconnaissance is (e.g., probability of effectual reconnaissance~\cite{huang2020dynamic}) or accurately determining whether the target node is real or fake ~\cite{bilinski2019you}). However, not much work considers the operating system versions or other node configurations for reconnaissance.

As defenders initiate honeypot placement at different locations throughout the attack graph~\cite{kiekintveld2015game,durkota2015optimal,milani2020harnessing}, the attacker needs to identify whether the next target is a real node or honeypot. In most game-theoretic models the authors assume honeypots are a controlled environment where they can detect if an attacker compromises these nodes, thus leading to attacker engagement games~\cite{huang2019adaptive}. 
Adversarial reconnaissance is described as a variety of scanning techniques for determining the network's topology, node configurations, and the location of nodes~\cite{kelly2019adversarially} or application-oriented information (e.g., database server information~\cite{ferguson2019game}).

\subsubsection{Active vs. Passive Reconnaissance}
Most of the literature discusses in detail the defender's perspective while modeling the attacker's recon behavior as simple scanning or probing. For example, Crouse et al. describe that the
attacker can attempt $k$ reconnaissance actions serially (one at a time) or in parallel (all $k$ at once)~\cite{crouse2015probabilistic}. A full scan refers to $k=n$ where $n$ is the number of nodes in the network. The attacker's success depends on the result of the scan, and if the attacker can identify at least one vulnerable host. 
Attackers can also perform passive reconnaissance by analyzing captured network packets which leads to the
defender's strategies of traffic obfuscation (e.g., honey traffic~\cite{anjum2020optimizing}).


\subsubsection{Reconnaissance Observation Models}

Most deception games do not have a detailed attacker observation model, with most using simple models that assume high-level information that is observed with precision.  For example, in signaling games~\cite{pawlick2018modeling,pawlick2015deception,durkota2015optimal}, reconnaissance is simply defined by the attacker perception of the received signal. However, some game models consider the attacker perceiving partial or full observation from the environment. For example, 
Huang et al. formed a multi-stage dynamic game to deceive Advanced Persistent Threats in cyber-physical systems where the authors talked about different payoffs if there is an effectual reconnaissance or not~\cite{huang2020dynamic}. In this game, both players (stealthy attacker and proactive defender) have incomplete information about each other.

\subsection{Reconnaissance in Non-Game Theoretic Deception Models}\label{subsec:non-game-theory-models}
Very few works capture adversarial reconnaissance using belief update models in a network where at least one node has been compromised. Some previous work~\cite{jajodia2017probabilistic,sugrim2018measuring} describes concepts of reconnaissance while considering node configurations and belief update approaches (e.g., Bayesian inference, tightening probability intervals).

Jajodia et al. argued that attackers could map system configurations (e.g., type of operating systems, applications, or services) for a particular node in the network~\cite{jajodia2017probabilistic}. The authors propose a belief state model that considers an interval of probabilities for specific configurations and then tightens the interval over time. The authored assume possible true configuration based on the probability interval, sometimes leading to uncertain decisions. Moreover, the weight, initial belief lower bound, and upper bound are user-defined and can eventually lead to different belief states. 

Sugrim et al. utilize Bayesian inference to update attacker's belief for an individual node property (e.g., IP address)~\cite{sugrim2018measuring}. The authors have shown that their model can capture existent or non-existent hosts in the environment. However, they did not consider node configurations (operating system, application, or services) as part of their model. Considering that the reconnaissance includes the node configuration scanning, it takes time to update belief considering each configuration as a separate node property. This procedure is a naive approach and can misbehave if different operating systems have similar feature values, i.e., operating system fingerprinting features (e.g., TTL, window size).

\paragraph{Comparison of Different Reconnaissance Models}
We studied the literature to find whether the existing works considered node configurations (installed OS, applications, or services) as part of reconnaissance, if the recon is active or passive, what are the target information, if authors considered model for observation and knowledge base, and if the authors considered a belief update model over a number of observations.  
Table~\ref{tab:lit_comparison} presents the comparison of reconnaissance approaches deception games and closely related non-game theoretic deception models. We find that most game-related works consider more simplistic models that do not define node configurations and belief update models. However we believe these are critical to finding better deception strategies.
\begin{table}[!ht]
\centering
\caption{Comparison of Existing Reconnaissance Models}
\label{tab:lit_comparison}

\begin{tabular}{|c|c|c|c|c|c|c|}
\hline
 Genre & \diagbox[width=8em]{Works}{Features} & \begin{tabular}[c]{@{}c@{}}Node\\Configuration \end{tabular} &  \begin{tabular}[c]{@{}c@{}}Type\end{tabular} & \begin{tabular}[c]{@{}c@{}}Target\\Info\end{tabular} & \begin{tabular}[c]{@{}c@{}}Observation\\/KB model\end{tabular} & \begin{tabular}[c]{@{}c@{}}Belief\\Model\end{tabular} \\ \hline\hline
\multirow{2}{*}{\begin{tabular}[c]{@{}c@{}}Recon\\Models\end{tabular}} & \begin{tabular}[c]{@{}c@{}}Jajodia et\\ al., 2017~\cite{jajodia2017probabilistic}\end{tabular} & \checkmark & \begin{tabular}[c]{@{}c@{}}Active/\\Passive\end{tabular} & OS/SW & \text{\sffamily X} & \begin{tabular}[c]{@{}c@{}}Node/\\Network\end{tabular} \\ \cline{2-7} 
 & \begin{tabular}[c]{@{}c@{}}Sugrim et\\ al., 2018~\cite{sugrim2018measuring}\end{tabular} & \text{\sffamily X}  & Active & IP & \text{\sffamily X} & Network \\ \hline
\multirow{8}{*}{\begin{tabular}[c]{@{}c@{}}Game\\Models\end{tabular}}
 & \begin{tabular}[c]{@{}c@{}}Sengupta et\\ al., 2018~\cite{sengupta2018moving}\end{tabular}  & \text{\sffamily X} & \begin{tabular}[c]{@{}c@{}}N/A\end{tabular} & OS/SW & \text{\sffamily X} & \text{\sffamily X} \\ \cline{2-7} 
 & \begin{tabular}[c]{@{}c@{}}Milani et\\ al., 2020~\cite{milani2020harnessing}\end{tabular} & \text{\sffamily X}  & N/A & \begin{tabular}[c]{@{}c@{}}Node\\reward\end{tabular} & \text{\sffamily X} & \text{\sffamily X} \\ \cline{2-7} 
 & \begin{tabular}[c]{@{}c@{}}Anwar et\\ al., 2018~\cite{anwar2020game}\end{tabular} &  \text{\sffamily X} & N/A & \begin{tabular}[c]{@{}c@{}}Attack\\Graph\end{tabular} & \text{\sffamily X} & \text{\sffamily X} \\ \cline{2-7} 
 & \begin{tabular}[c]{@{}c@{}}Huang et\\ al., 2019~\cite{huang2019adaptive}\end{tabular} & \text{\sffamily X}  & N/A & \begin{tabular}[c]{@{}c@{}}Node/\\Honeypot\end{tabular} & \text{\sffamily X} & \text{\sffamily X} \\ \cline{2-7} 
 & \begin{tabular}[c]{@{}c@{}}Pawlick et\\ al., 2018~\cite{pawlick2018modeling}\end{tabular} & \text{\sffamily X} & N/A & \begin{tabular}[c]{@{}c@{}}Node/\\Honeypot\end{tabular} & \text{\sffamily X} & \text{\sffamily X} \\  \hline
\end{tabular}
\end{table}




\section{Adversarial Reconnaissance Belief Model} \label{sec:passive_recon_model}

We now introduce a model for updating beliefs during adversarial reconnaissance that used an explicit knowledge base and Bayesian updates (Section~\ref{subsec:model}). Next, we illustrate this model on a real data set (Section~\ref{subsec:numerical_case}).

\subsection{Network Configurations and Adversary Beliefs}\label{subsec:model}

\subsubsection{Node Configuration}
A defenders' network $T$ is a set of nodes,
where $N \in T$ denotes a node. Based on the real configuration of a node, an adversary can identify vulnerabilities of the installed OS and software.
We define a node's real configuration as $N = \langle A_N, o_N, S_N \rangle$, where
$A_N$ is the set of addresses (IP, MAC addresses), $o_N$ is the installed OS, and $S_N$ is the set of installed software, applications, or services.
For example, a node could have the configuration $N_1 =  \langle$192.168.1.5,  Red Hat 7, Apache HTTP 2.4.0$\rangle$.


\subsubsection{Adversary Beliefs}

%
We consider an attacker's belief to be any valid configuration of a particular node.
In the network $ T $, the set of all possible operating systems is $ O $, where a particular operating system is $o \in O $, and the set of all possible software $ S $, with specific $s \in S$.
The set of all possible configurations for a node can be expressed as
$\calW_{N} = O \times \calP(S)$, where $\calP(S)$ is the power-set of $S$.
The attacker's belief is a distribution over possible specific configurations, i.e., $\beta(\textit{configuration}) \rightarrow [0,1]$.

The attacker considers each possible configuration in $\calW_{N}$ for a specific node.  If $\beta_N(\omega)$ is a belief for a node $N$, where $\omega \in \calW_{N}$, the probability value of $\beta_N$ can be defined as $0 \leq \beta_N(\omega) \leq 1$, and $\sum_{\omega \in \calW_{N}} \beta_N(\omega) = 1$.
Now, all possible configurations of a network $T$ can be expressed as the Cartesian product of configurations of individual nodes: $\calW_T = \calW_{N_1} \times \calW_{N_2} \times ... \times \calW_{N_{|T|}}$.

The attacker considers each possible combination of node configurations in $\calW_{T}$ as a configuration for the whole network. We denote $\beta_T(\tau)$ as a specific belief for the whole network $T$, where $\tau \in \calW_{T}$. Here, the probability value of $\beta_T$ is defined by $0 \leq \beta_T(\tau) \leq 1$ and $\sum_{\omega \in \calW_{T}} \beta_T(\tau) = 1$.

\subsubsection{Observation and Knowledge Base}

In order to update beliefs the attacker needs a knowledge base that relates observations to the underlying network configurations. 
In practice, the attacker can calculate the probabilities of an observation given a particular configuration (an OS or a software) from data. We denote a set of observations as $\calE$ for node $N$, where each observation $\Theta \in \calE$, and $\Omega \in \Theta$  represents as an observation feature that has a value $v$, i.e., $\Omega = v$.  

If the set of OS fingerprinting features is $\calF_O(\Theta)$, and the set of software fingerprinting features is $\calF_S(\Theta)$ for a node $N$, the probability of an observation given a configuration can be denoted by
$\Pr(\Theta \,|\, o_N = o) = \Pr( \bigcap_{\Omega \in \calF_O(\Theta)} \Omega = v  \,|\, o_N = o)  = [0,1]$
for an installed operating system $o \in O$ and $\Pr(\Theta \,|\, s_N = s) = \Pr( \bigcap_{\Omega \in \calF_S(\Theta)} \Omega=v  \,|\, s_N = s) = [0,1]$ for an installed software $s \in S$.
Now, the attacker can specify the probability of an observation given a particular configuration $\omega \in \calW_N$ as
$\Pr(\Theta \,|\, \omega) = \Pr(\Theta \,|\, o_N = o, S_N = S) 
= \Pr( \bigcap_{\Omega \in \calF_O(\Theta)} \Omega = v  \,|\, o_N = o) \cdot \prod_{s \in S} \Pr( \bigcap_{\Omega \in \calF_S(\Theta)} \Omega = v  \,|\, s_N=s)$.

\begin{algorithm}[H]
\DontPrintSemicolon
  
\tcc{KB $\rightarrow$ Knowledge Base}
\tcc{$KB^{o \in O} \leftarrow $ KB of $\Pr(\Theta \,|\, o_N = o)$ for $o \in O$}
\tcc{$KB^{s \in S} \leftarrow $ KB of $\Pr(\Theta \,|\, s_N = s)$ for $s \in S$}

\KwInput{$\calW_N, \calE, KB^{o \in O}, KB^{s \in S}$ }
\KwOutput{$\omega_{MAP}$}

calculate prior belief $\beta_N(\omega)= \Pr(o_N = o, S_N = S) = \frac{1}{|\calW_N|}$ for $\omega \in \calW_N$\;
calculate $\Pr(\Theta \,|\,  \omega)$ for $\omega \in \calW_N$\;
    
    \ForEach{ observation $\Theta \in \calE$}    
        {   
            \ForEach{ configuration $\omega \in \calW_N$}    
                { 
                    update probability of $\omega \in \calW_N$ given $\Theta$ as $\Pr(\omega \,|\, \Theta ) = \Pr(o_N = o, S_N = S \,|\, \Theta) =\frac{\Pr(\Theta \,|\, o_N = o, S_N = S) \cdot \Pr(o_N = o, S_N = S)}{\sum_{\omega^* \in \calW_N} \big [ \Pr(\Theta \,|\, o_N = o^*,\, S_N = S^*) \cdot \Pr(o_N = o^*, \,S_N = S^*) \big ]} $ \;
                    update $\Pr(o_N = o, S_N = S) = \Pr(\omega \,|\, \Theta )$\;
                }
        }

\caption{Attacker's Belief Update For a Node}
\label{algo:belief_update}
\end{algorithm}

\subsubsection{Adversary Belief Updates}
Adversaries have an initial prior belief over possible configurations. Before making any specific observations, attackers may (or may not) believe that all configurations are equally likely. If the belief is uniform, we get $\sum_{\omega \in \calW_{N}} \beta_N(\omega) = 1$ and $\beta_N(\omega_1) = \beta_N(\omega_2) = ... = \beta_N(\omega_{|\calW_N|}) = \frac{1}{|\calW_N|}$.
Similarly, for a whole network $T$, a uniform prior belief would specify that all initial beliefs of node configurations are equal: $\sum_{\tau \in \calW_{T}} \beta_T(\tau) = 1$ and $\beta_T(\tau_1) = \beta_T(\tau_2) = ... = \beta_T(\tau_{|\calW_T|}) = \frac{1}{|\calW_T|}$.

The installed software in a node typically depends on the installed operating system due to platform dependency. For instance, particular software (e.g., Paint.NET, MS Security Essential) can be installed in Windows OS and has no package distributions for other platforms. Similarly, some software packages depends on each other, e.g., some software may require the Java run-time environment. The attacker needs to update each belief for each node $N \in T$, and may use these dependencies. Using the \emph{chain rule}, the attacker's belief is updated as: 
$\Pr(o_N = o, S_N = S) = \Pr(s_n \,|\, s_{n-1},...,s_1,o) \cdot \Pr(s_{n-1},...,s_1,o)$, where $o \in O, s \in S, n = |S|$.

An attacker updates beliefs about the network by analyzing observations of network data that can (partially) reveal configuration information.  
For each node the attacker updates $\omega \in \calW_N$ after each observation $\Theta \in \calE$ using Bayesian updates. The prior probability for the first step is $\Pr(o_N = o, S_N = S) = \beta_N(o, S) = \frac{1}{|\calW_N|}$
and for the later steps is
$\Pr(o_N = o, S_N = S) = \Pr(\omega \,|\, \Theta )$.


The attacker can use \emph{Maximum a posteriori} (MAP) to find the most likely final configuration of a node; for $\omega \in \calW_N$, we  define MAP as 
$\omega_{MAP} = \operatorname{argmax}_{\omega^* \in \calW_N} \, \Pr(\omega^* \,|\, \Theta )$.
Algorithm~(\ref{algo:belief_update}) details the belief updates.

\begin{figure}[t]
\pgfplotstableread[col sep=comma,]{prob.csv}\datatable
\pgfplotstableread[col sep=comma,]{num.csv}\table
\centering

\begin{minipage}{.48\linewidth}
\centering
{
\begin{tikzpicture}
\begin{axis}[
  width=\textwidth,
  height=0.8\textwidth,
  ymax=1.15,
  ymin=0, 
  legend style={at={(0.95,0.1)},
  anchor=south east}, 
  xtick=data, 
  xticklabels from table={\datatable}{Iter}, 
  font=\scriptsize,
  x tick label style={rotate=35, anchor=east, yshift=-4pt}, 
  xlabel={Node IP},
  ylabel={Probability}]

  \addplot[only marks,mark=*,mark options={fill=green}] table[x expr=\coordindex,y index=2,col sep=comma] {prob.csv};   
\addlegendentry{Probability}

\addplot[only marks,mark=triangle,mark options={fill=cyan,scale=2}] table[x expr=\coordindex,y index=1,col sep=comma] {prob.csv}; 
\addlegendentry{Ground Truth}
\end{axis}
\end{tikzpicture}}
    \caption{Belief Probability vs. Ground Truth}
    \label{fig:ground_truth}
\end{minipage} \hspace{.6em}
\begin{minipage}{.48\linewidth}
\centering
{
\begin{tikzpicture}
\begin{axis}[
    SmallBarPlot,
    width=\textwidth,
    height=0.8\textwidth,
    font=\scriptsize,
    ymin = 0,
    xtick=data,
    xticklabels from table={\table}{IP},
    x tick label style={rotate=35, anchor=east, yshift=-4pt},
    xlabel={Node IP},
    ylabel={Number of Observations}]
    \addplot [BlueBars2] table [x expr=\coordindex, y={num}]{\table};
\end{axis}
\end{tikzpicture}}
    \caption{Number of Observations Necessary}
    \label{fig:prob_eval}
\end{minipage}

\caption{(a) Evaluating probability of each node configuration with ground truth; (b) number of observations required to achieve at least $0.9$ probability on the ground truth.}
\label{fig:subfig_2}
\end{figure}

\subsection{Numerical Case Study} \label{subsec:numerical_case}

Now we present an initial case study to show the operation and feasibility of this type of detailed belief model for capturing the evolution of attacker beliefs using a real data set. 
 We used the CIC-IDS2017 dataset \cite{sharafaldin2018toward} to demonstrate our belief update model, and we show that final probability of the real configuration is close to the ground truth after a reasonable number of observations. We separated the dataset to create an OS fingerprinting knowledge base (using OS features such as TTL, Window Size, etc. extracted from one PCAP file), and then to show the evolution of the attacker's belief of installed OS in individual nodes (using another PCAP file as the attacker's observations).

\begin{figure}[t]
    \centering
\pgfplotstableread[col sep=comma,]{192_19_node_confT.csv}\datatable
\begin{tikzpicture}
\begin{axis}[
    width=0.95\textwidth,
    height=0.45\textwidth,
    xtick=data,
    xticklabels from table={\datatable}{configurations},
    x tick label style={font=\normalsize, rotate=35, 
    anchor=east
    },
    legend style={at={(0.98,0.43)},anchor=south east},
    ylabel={Probability}]
    
    \addplot [mark=o, black!70 ] table [x expr=\coordindex, y={win}]{\datatable};
    \addlegendentry{Windows}
    
    \addplot [mark=o, blue!70] table [x expr=\coordindex, y={ubuntu}]{\datatable};
    \addlegendentry{Ubuntu}
    
    \addplot [mark=o, red!70] table [x expr=\coordindex, y={mac}]{\datatable};
    \addlegendentry{macOS}
    
\end{axis}
\end{tikzpicture}
\caption{Belief update for a node (IP: 192.168.10.19) over number of observations.}
\label{fig:node_belief_update}
\end{figure}
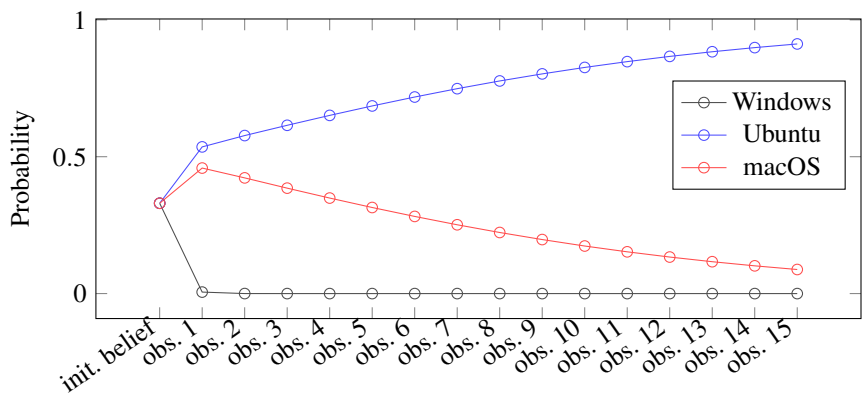

\subsubsection{Ground Truth vs. Calculated Probability}
Figure~\ref{fig:ground_truth} shows the probability values that we obtained from our model with respect to the ground truth. We observe that the predictions of our model reliably converged to the ground truth. 
The model achieved a belief close to $1$ with a reasonable number of observations. Figure~\ref{fig:prob_eval} shows the number of observations required to achieve a configuration probability of $0.9$ or higher that matches the true configurations. Critically, we also see that the details of the belief updates matter; in some cases the beliefs converge quickly, but in other cases (e.g., Unix-like systems that have closer feature values) it may require many ($15$ in this example) data points to converge. 





\subsubsection{Node Belief Update}
Figure~\ref{fig:node_belief_update} shows the belief update for a node ($192.168.10.19$) where the installed operating system is Ubuntu. The plot shows the gradual belief updates after calculating the probability for each initial configurations (the three main operating systems in the dataset) after each observation. We see gradual trends for ``Ubuntu'' and ``macOS'' since both are similar operating systems.

\paragraph{Remarks}
The primary focus of this numerical case study is to show how the attacker's belief evolves with increasing observations. Most models in the deception literature fail to capture the details of this evolution or the uncertainty that arises from realistic observations of network data (Figure~\ref{fig:node_belief_update}). This could lead to many more options for defenders to optimize deceptions strategies based on a detailed understanding of attacker beliefs. The tradeoff is that this requires more data, more complex models, and potentially increased computation time to evaluate these models to optimize the defensive strategies. Developing more sophisticated methods for all of these presents great opportunities for additional research.

\section{Discussion}
In this section, we discuss learned lessons and highlight key findings of the chapter.

\subsection{Analyzing Reconnaissance to Improve Security}
Security awareness and best practices can mitigate the adversarial reconnaissance at a good level. As reconnaissance is the prerequisite part of any further progress made by adversaries, defenders are required to take necessary measures that prevent or mitigate the effect of recon techniques.

\vspace{3mm}
Table~\ref{tab: defensive_measurements} presents which defensive measures can counter particular reconnaissance techniques. In this table, we see that \emph{Social Engineering}, \emph{Network Scanning}, and \emph{Local Discovery} techniques can be detected and thus be prevented with effective measures. \emph{Deception} and \emph{MTD}-based techniques can mitigate adversarial scanning and local discovery techniques. Above all, security awareness (through employee training programs) and known best practices can effectively mitigate adversarial techniques to a diverse extent.

\begin{table}[!ht]
\centering
\caption{Defensive Measures against Reconnaissance Techniques}
\label{tab: defensive_measurements}
\begin{tabular}{|c|c|c|c|c|}
\hline
\diagbox[width=15em]{Measures}{Techniques}     & \begin{tabular}[c]{@{}c@{}}Target\\ Footprinting\end{tabular} & \begin{tabular}[c]{@{}c@{}}Social \\ Engineering\end{tabular} & \begin{tabular}[c]{@{}c@{}}Network\\ Scanning\end{tabular} & \begin{tabular}[c]{@{}c@{}}Local\\ Discovery\end{tabular} \\ \hline\hline
Reconnaissance Detection & \textbf{\textsf{X}} & \checkmark & \checkmark & \checkmark \\\hline
Cyber Deception/MTD   &  \textbf{\textsf{X}}   &  \textbf{\textsf{X}} &  \checkmark & \checkmark \\ \hline
Security Awareness and Best Practices & \checkmark  & \checkmark &  \checkmark &  \checkmark \\ \hline
\end{tabular}
\end{table}

\subsection{Limitations of Existing Deception Games}\label{subsec:lit_limitation}


Most of the related work in deception games has not discussed reconnaissance in detail while building game models in a deceptive network. For example,
Zhang et al. considered bypassing the scanning detection tool while forming a dynamic Markov differential game model~\cite{zhang2020moving}. Many articles including game models based on honeypot~\cite{kiekintveld2015game,durkota2015optimal} or IDS~\cite{sengupta2018moving} placement, modifying attack graphs by hiding or adding real or fake nodes~\cite{milani2020harnessing,anwar2020game}, attacker engagement~\cite{huang2019adaptive} based signaling games~\cite{pawlick2018modeling,pawlick2015deception,durkota2015optimal} and Stackelberg games~\cite{feng2017stackelberg,durkota2015optimal,sengupta2018moving} developed the deception games without discussing or including a detailed reconnaissance model. Specifically in signaling games, it is unclear how defender-generated signals can affect attacker reconnaissance for identifying actual nodes or honeypots. Game models including Feng et al.~\cite{feng2017stackelberg} (assuming the attacker can attack any valid state of all resources, attacking cost is identical for all states, and choosing to attack is optimal for the attacker) and Horak et al.~\cite{horak2017manipulating} (defining attacker actions as taking control of nodes and escalating privileges) have not defined how the adversarial reconnaissance is performed in their game models. Kiekintveld et al. built a honeypot-based attack graph model to deceive the attacker, where they defined the attacker's action as selecting the next target based on probability without discussing how reconnaissance is modeled to set a value for explored nodes~\cite{kiekintveld2015game}. Many papers (e.g., Huang et al.) also present reconnaissance as a binary set (effectual or not)~\cite{huang2020dynamic}.

The existing literature discusses the use of specific reconnaissance tools~\cite{poschinger2020openmtd,kiekintveld2015game,durkota2015optimal,albanese2015deception} but they do not consider whether the reconnaissance is active or passive, or details like how many times scanning is performed before forming final beliefs about network configurations.
Many works consider recon as a simple step where the attacker can see whatever configurations have already been changed by the defender.
Several papers~\cite{durkota2015optimal,fraunholz2018catch,kiekintveld2015game} mention the use of tools implying that the authors have not developed a model for a knowledge base rather than directly utilizing the tools' database which limits a clear understanding of how deception can affect the attacker, and the model of attacker observations is similarly vague.

\subsection{Importance of a Reconnaissance Model}
Deception strategies are developed according to the techniques used by the attackers to collect information. As artificial intelligence and game-theoretic models can effectively be used for strengthening and optimizing cyber defense, researchers need to work on reconnaissance belief update models first. Cyber reconnaissance is performed over a long period within the enterprise network, which can lead to further damage to organization assets. Future AI and game-theoretic models can effectively consider reconnaissance as a belief update model and build deceptive strategies around it, which can lead to further realistic security-based mathematical models.

\section{Conclusion and Future Works}\label{sec:conclusion}

We survey existing reconnaissance models and find that most existing models use simplistic approaches for modeling what an adversary knows about the network that do not consider the complexities of real network observations, the uncertainties in realistic beliefs, or the evolution of beliefs with increasing amounts of data. We introduce a model that can solve these issues by reasoning in detail about the attackers beliefs using Bayesian updates based on a realistic knowledge base. 

There are many issues that this raises for future work, including the scalability challenges of trying to reason about large networks using exact (or approximate) Bayesian reasoning. There are also many issues that need further investigation in creating knowledge bases from data, including how to select relevant features, how to deal with very large feature spaces to build useful knowledge bases, etc. In addition, our current adversary model does not explicitly consider deception; we plan to extend this model to consider the effects of deception on belief formation in future work.

\bibliographystyle{ACM-Reference-Format}
\bibliography{main}

\end{document}